\documentclass[10pt,aps,prd,nofootinbib,superscriptaddress,twocolumn,preprintnumbers,balancelastpage]{revtex4}

\usepackage{amssymb,amsmath,latexsym,graphics, graphicx,epsfig,multirow,comment,appendix,feyn,slashed,color,afterpage,multirow,makecell} 

\usepackage[colorlinks=true
,urlcolor=blue
,anchorcolor=blue
,citecolor=blue
,filecolor=blue
,linkcolor=blue
,menucolor=blue
,linktocpage=true
,pdfproducer=medialab
,pdfa=true
]{hyperref}

\newcommand {\be} {\begin {equation}}
\newcommand {\ee} {\end {equation}}

\newcommand {\bes} {\begin {equation*}}
\newcommand {\ees} {\end {equation*}}

\def\CO{{\cal O}}

\newcommand{\beq}{\begin{equation}}
\newcommand{\eeq}{\end{equation}}

\newcommand{\kev}{\,\textrm{keV}}
\newcommand{\mev}{\,\textrm{MeV}}

\newcommand{\khz}{\,\textrm{kHz}}

\newcommand{\alpNP}{\alpha_{\rm NP}}
\newcommand{\Uhat}{ \overrightarrow{m\mu}}

\newcommand{\mnvec}{\overrightarrow{m\nu}}
\newcommand{\mrvec}{ \overrightarrow{m \delta \langle r^2 \rangle}}

\begin{document}

\title{Probing new light force-mediators by isotope shift spectroscopy}

\preprint{DESY 17-055, FERMILAB-PUB-17-077-T, LAPTh-009/17, MIT-CTP-4898}

\author{Julian C. Berengut}
\email{julian.berengut@unsw.edu.au}
\affiliation{School of Physics, University of New South Wales, Sydney, New South Wales 2052, Australia}

\author{Dmitry Budker}
\email{budker@uni-mainz.de}
\affiliation{Helmholtz-Institut Mainz, Johannes Gutenberg-Universit\"{a}t Mainz, 55128 Mainz, Germany}
\affiliation{Physics Department, University of California, Berkeley 94720-7300, USA}
\affiliation{Nuclear Science Division, Lawrence Berkeley National Laboratory, Berkeley, California 94720, USA}
\author{C\'edric Delaunay}
\email{cedric.delaunay@lapth.cnrs.fr}
\affiliation{Laboratoire d'Annecy-le-Vieux de Physique Th\'eorique LAPTh, CNRS -- Universit\'e Savoie Mont Blanc, BP 110, F-74941 Annecy-le-Vieux, France}
\author{Victor~V.~Flambaum}
\email{v.flambaum@unsw.edu.au}
\affiliation{School of Physics, University of New South Wales, Sydney, New South Wales 2052, Australia}
\author{Claudia Frugiuele}
\email{claudia.frugiuele@weizmann.ac.il}
\affiliation{Department of Particle Physics and Astrophysics, Weizmann Institute of Science, Rehovot 7610001, Israel}
\author{Elina Fuchs}
\email{elina.fuchs@weizmann.ac.il}
\affiliation{Department of Particle Physics and Astrophysics, Weizmann Institute of Science, Rehovot 7610001, Israel}
\author{Christophe Grojean}
\email{christophe.grojean@desy.de}
\affiliation{DESY, D-22607 Hamburg, Germany}
\affiliation{Institut f\"ur Physik, Humboldt-Universit\"at zu Berlin, D-12489 Berlin, Germany}
\author{Roni Harnik}
\email{roni@fnal.gov}
\affiliation{Theoretical Physics Department, Fermi National Accelerator Laboratory, Batavia, IL 60510 USA}
\author{Roee Ozeri}
\email{roee.ozeri@weizmann.ac.il}
\affiliation{Department of Physics of Complex Systems, Weizmann Institute of Science, Rehovot 7610001, Israel}
\author{Gilad Perez}
\email{gilad.perez@weizmann.ac.il}
\affiliation{Department of Particle Physics and Astrophysics, Weizmann Institute of Science, Rehovot 7610001, Israel}
\author{Yotam Soreq}
\email{soreqy@mit.edu}
\affiliation{Center for Theoretical Physics, Massachusetts Institute of Technology, Cambridge, MA 02139}


\begin{abstract}
\begin{center}
{\bf Abstract}
\end{center}
In this Letter we explore the potential of probing new light force-carriers, with spin-independent couplings to the electron and the neutron, using precision isotope shift spectroscopy. We develop a formalism to interpret linear King plots as bounds on new physics with minimal theory inputs. 
We focus only on bounding the new physics contributions that can be calculated independently of the Standard Model nuclear effects.
We apply our method to existing Ca$^+$ data and project its sensitivity to possibly existing new bosons using narrow transitions in other atoms and ions (specifically, Sr and Yb). Future measurements are expected to improve the relative precision by five orders of magnitude, and can potentially lead to an unprecedented sensitivity for bosons within the 10\,keV to 10\,MeV mass range. 
\end{abstract}
\maketitle

\section{Introduction} 
\label{sec:intro}

The Standard Model of particle physics~(SM) successfully describes multiple observations up to the TeV scale, and is theoretically consistent up to a much higher energy.  
However, the SM cannot be a complete description of Nature. For example, it lacks a viable dark matter candidate and can neither explain the observed matter-antimatter asymmetry of our Universe nor neutrino oscillations. In addition, the SM suffers from hierarchy issues both in the Higgs sector and  the fermionic sector. These experimental observations require new physics~(NP) beyond the SM, however, none of these observations point towards a specific new theory or energy scale. 

The quest for NP is pursued in multiple directions. Current efforts with colliders such as the LHC form the energy frontier, probing directly the TeV energy scale. Other accelerators, such as $B$-factories, NA62 and neutrino experiments, form the intensity frontier that broadly probes the MeV--GeV scale. 
Atomic physics tabletop experiments form a third frontier of precision measurements (see e.g.: \cite{Baron:2013eja,PhysRevLett.100.120801,bouchendira2011new,wood1997measurement,guena2005measurement}, for a review see~\cite{Budker:2017ReviewInPrep,Karshenboim20051,Ginges:2003qt}) 
where sub-MeV physics can be efficiently tested. It is interesting to note that NP that may account for the hierarchy issues could be new light scalars that couple to matter fields~\cite{Kim:1986ax, Graham:2015cka, Gupta:2015uea, Flacke:2016szy, Feng:1997tn,Wilczek:1982rv,Gelmini:1982zz}.
To convert the high precision offered by atomic and molecular spectroscopy into sensitivity to fundamental new physics, one either has to acquire similar theoretical accuracy of atomic structure or alternatively seek for unique observables that are insensitive to theoretical uncertainties.

\begin{figure}[t]
\begin{center}
\includegraphics[width=\columnwidth]{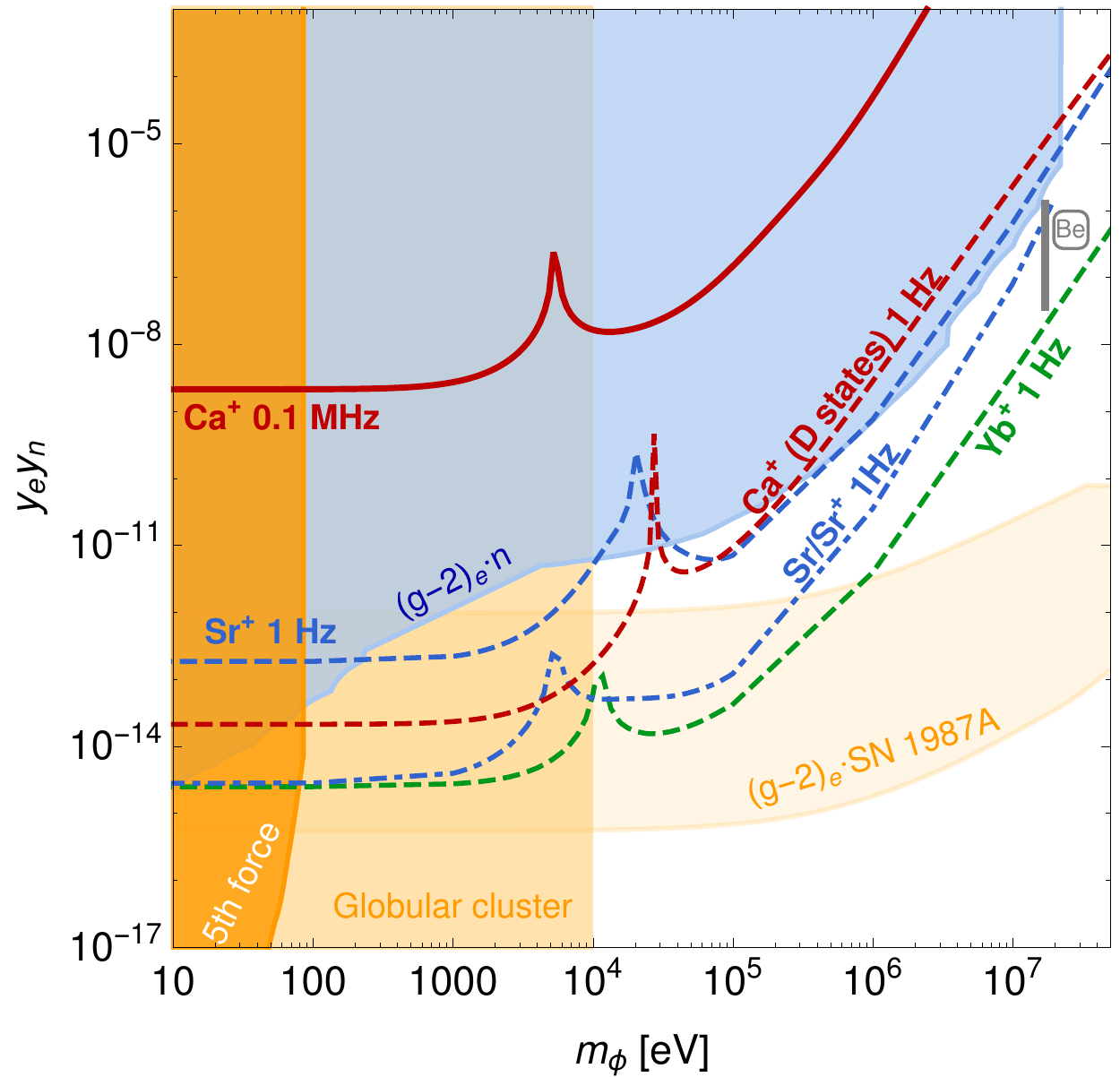}
\caption{
Limits on the electron and neutron couplings ($y_e y_n$) of the new boson of mass $m_{\phi}$ (for the experimental accuracies $\sigma_i$ specified in the labels).
Constraint from existing IS data: Ca$^+$ (397\,nm vs. 866\,nm~\cite{PhysRevLett.115.053003}, solid red line).
IS projections (dashed lines) for Ca$^+$ ($S\to D$ transitions), Sr$^+$, Sr/Sr$^+$,  and Yb$^+$. For comparison, existing constraints from other experiments (shaded areas): fifth force~\cite{Bordag:2001qi,bordag2009advances} (dark orange), 
 $(g-2)_e$~\cite{Olive:2016xmw,Hanneke:2010au} 
combined with neutron scattering~\cite{Barbieri:1975xy,Leeb:1992qf,Nesvizhevsky:2007by,Pokotilovski:2006up} (light blue)
or SN1987A~\cite{Raffelt:2012sp} (light orange), and from star cooling in globular clusters~\cite{Yao:2006px,Grifols:1986fc,Grifols:1988fv} (orange). 
The gray line at $17\mev$ indicates the $y_ey_n$  values required to accommodate the Be anomaly~\cite{Feng:2016jff,Feng:2016ysn}.
}
\label{fig:bounds}
\end{center}
\end{figure}

In this paper we show that precision isotope shift~(IS) spectroscopy may probe spin-independent couplings of light boson fields to electrons and neutrons. 
The idea is to extract constraints from bounds on nonlinearities in a King plot comparison~\cite{King:63} of isotope shifts of two narrow transitions~\cite{Delaunay:2016brc}. 
We develop a new formalism to interpret these measurements in the context of searching for new light force carriers and propose several elements and transitions that can be used for such analyses. 
We recast existing measurements into bounds and provide an estimation for the sensitivity of future measurements, see Fig.~\ref{fig:bounds}.
The validity of our method to bound NP does not rely on the knowledge of the SM contributions to King plot nonlinearites. 
Its constraining power, however, is limited by the size of the observed nonlinearities.
In case that Kings linearity is established, at the current state-of-the-art experimental precision, and baring cancellation between the SM and NP contributions, world-record sensitivity in a certain mass range will be achieved. 

\section{Factorization of Nuclear and Atomic Effects in Isotope Shifts} 
\label{sec:NLKPNP}

We now discuss the scaling and factorization properties of IS which we use to probe NP in this work. Consider an atomic transition, denoted by $i,$  between narrow atomic states. The difference in the transition frequency $\nu$ comparing the isotopes $A$ and $A'$ is the IS, 
\begin{equation}
	\nu_i^{AA'}\equiv \nu_i^{A}-\nu_i^{A'}\,.
\end{equation}
At leading order~(LO) the IS receives contributions from two sources, mass shift (MS) and field shift (FS). Mass shift arises due to a correction to the kinetic energy of atomic electrons due to the motion of the nucleus. For independent electrons, this is just replacing $m_e $ by the reduced mass but if electrons are correlated, this could be orders of magnitude larger. Field shift originates from different contact interactions between electrons and nuclei in isotopes. Putting these two leading contributions together, the IS can be phenomenologically written as
\begin{align}
	\label{eq:ISLO}
	\nu_i^{AA'} 
=	K_i\, \mu_{AA'} + F_i\, \delta\langle r^2 \rangle_{AA'} +\ldots \, ,
\end{align}
where the two terms represent MS and FS respectively~\cite{King:63,King:13}. We define $\mu_{AA'} \equiv m^{-1}_A - m^{-1}_{A'} $, where $m_A$ and $m_{A^{\prime}}$ are the masses of isotopes~$A$ and~$A^{\prime}$. 

The quantity   $\delta\langle r^2 \rangle_{AA'}$ is dominated by the difference in the mean squared charge radii of the two nuclei but can include other contact interactions. Both $\mu_{AA'}$ and  $\delta\langle r^2 \rangle_{AA'}$  are purely nuclear quantities that \emph{do not} depend on the electronic transition $i$. Note, however, that $\mu_{AA'}$ is known with high precision, whereas $\delta \langle r^2 \rangle_{AA'}$ is known only to a limited accuracy. The parameters $K_i$ and $F_i$ are isotope-independent, transition-dependent coefficients of the MS and FS, and their precise values are unnecessary in the observable we construct. 
Each term of Eq.~\eqref{eq:ISLO} is a product of a purely nuclear quantity and a purely electronic quantity, resulting in the factorization of nuclear and electronic dependence. This is known as LO factorization.

Given two electronic transitions, $i=1,2$, one can eliminate the uncertain $\delta\langle r^2 \rangle_{AA'}$ giving a relation between the isotope shifts $\nu_1^{AA'}$ and $\nu_2^{AA'}$. In terms of the modified IS\footnote{Below we will adopt the notation of adding an $m$ to ``modified'' (i.e. normalized by $\mu_{AA'}$) quantities, such as $m\delta \langle r^2 \rangle_{AA'} \equiv \delta \langle r^2 \rangle_{AA'} /\mu_{AA'}$.}, $m\nu_i^{AA'}\equiv \nu_i^{AA'}/\mu_{AA'}$, this relation is,
\begin{align}
	\label{eq:modKing}
	m\nu_2^{AA'}  \!=\! 	
	K_{21} \!+\! F_{21} m\nu_1^{AA'} \! \,,
\end{align}
with $F_{21}\equiv F_2/F_1$, and  $K_{21}\equiv K_2-F_{21}K_1$. 

Equation~\eqref{eq:modKing} reveals a linear relation between $m\nu_1$ and $m\nu_2$, giving rise to a straight line in the so-called King plot of $m\nu_2$ vs $m\nu_1$ ~\cite{King:63}. 
It is important to stress that the linearity of this equation holds regardless of the precise values of the $K_i$ and $F_i$ electronic parameters. Testing linearity necessitates at least three independent isotope pairs in two transitions, which constitutes a purely data driven test of LO factorization. 

The formulae in our treatment of NP will be simplified greatly by introducing a geometrical description of LO factorization. 
As we will now explain, King linearity is equivalent to coplanarity of vectors. 
For each transition $i$, we can form  a vector 
\begin{equation}
	\label{eq:mnuvec}
	\mnvec_i \equiv \left( m \nu_i^{AA'_1},m \nu_i^{AA'_2}, m\nu_i^{AA'_3} \right)\,.
\end{equation}
The nuclear parameters of field and mass shift, $\mu_{AA'}$ and $\delta \langle r^2 \rangle_{AA'}$ can also be written as vectors $\Uhat$ and  $\mrvec$ in the same space (notice that $\Uhat  \equiv  \left({1,1,1}\right)$) and hence Eq.~(\ref{eq:ISLO}) becomes
\begin{equation}
	\label{eq:nuAApVec}
	\mnvec_i = K_i\, \Uhat + F_i\, \mrvec.
\end{equation}
In this language LO factorization implies the following qualitative statement: any vector of reduced isotope shifts, $\mnvec_i$, must lie in the plane that is defined by  $\Uhat$ and $\mrvec$, as illustrated in the cartoon in the left panel of Fig.~\ref{fig:plane}.  

Note that, because the direction of $\mrvec$ in this space is uncertain, theory does not tell us in which direction this plane is oriented. 
However, by measuring two IS vectors, $  \mnvec_1$ and  $\mnvec_2$, we can test this statement by asking whether the three vectors $\mnvec_1$, $\mnvec_2$, and $\Uhat$ are coplanar. The coplanarity of these vectors corresponds to King linearity as we can see by
rewriting Eq.~(\ref{eq:modKing}) in vectorial form $\mnvec_2 =K_{21} \Uhat \!+\! F_{21} \mnvec_1$. 
Like King linearity, coplanarity is a purely data driven test of LO factorization since it is independent of theoretical input. A change in $K_i$ and $F_i$ will merely change the direction of $\mnvec_1$ and $\mnvec_2$ \emph{within the plane}, but the qualitative statement of coplanarity remains.

In this vector language we can provide a compact expression for a nonlinearity measure, 
\begin{align}
	\label{eq:NL}
	{\rm NL} = \frac{1}{2}\left|( \mnvec_1 \times  \mnvec_2 ) \cdot \Uhat\right|	\, .
\end{align}
In terms of the King plot, NL is the area of the triangle spanned by the three points shown in Fig.~\ref{fig:NL}. Equivalently, in our geometrical picture it is the volume of the parallelepiped defined by $\mnvec_{1,2}$ and  $\Uhat$.
A given data set is considered linear if NL is smaller than its first-order propagated error $\sigma_{\rm NL} = \sqrt{\Sigma_k(\partial \textrm{NL}/\partial \CO_k)^2 \sigma_k^2}$ where the sum runs over all measured observables $\CO_k$ (modified frequency shifts and isotope masses) with standard deviations $\sigma_{k}$. 

\section{New Physics and violation of King linearity }\label{sec:NP}

We now include a NP contribution by adding a third, also factorized,  term to Eq.~(\ref{eq:ISLO}),
\begin{align}
	\label{eq:nuAAp}
	\nu_i^{AA'} 
=	K_i\, \mu_{AA'} + F_i\, \delta\langle r^2 \rangle_{AA'} + \alpNP X_i\, \gamma_{AA'}\, ,
\end{align}
namely  $X_i$ depends on the form of the new potential and on the electronic transition, while $\gamma_{AA'}$ depends only on the nuclear properties. The parameter $\alpNP$ is the NP coupling constant which we would like to probe. 

Let us first mention two cases of NP which we do not expect to be able to probe by testing King linearity. For short-range NP (shorter than the nuclear size), the electronic parameters  $X_i$ will be proportional to those of FS, $X_i\propto F_i$. In this case the NP term can be absorbed
by redefining $\delta\langle r^2 \rangle_{AA'}$. Also, if the new physics couples to electrons and nuclei according to their electric charge (such as the case of dark-photon~\cite{Holdom:1985ag}), $\gamma_{AA'}=0$. There may also be cases in which NP can accidentally be absorbed by redefining $F_i$. However, a long-range force with couplings not proportional to the electric charge (and barring an accidental cancellation) can be severely constrained by tests of King linearity.   

Equation~\eqref{eq:modKing} written in vectorial form becomes
\begin{align}
	\label{KingNL}
	\mnvec_2  \!=\! 	
	K_{21} \Uhat \!+\! F_{21}  \mnvec_1  \! +\! \alpNP  \vec h X_1 \left( X_{21} \!-\! F_{21} \right)\,,
\end{align}
where $ \vec h $ is the NP vector in reduced frequency units, that is $h_{AA'}\equiv \gamma_{AA'}/\mu_{AA'}$ and $X_{21}\equiv X_2/X_1$.
One can see that NP can lead to a deviation from coplanarity if and only if 
(i)~the new force is not short-range,  $X_{21} \!\ne\!  F_{21}$;
(ii)~$\vec h$ is not aligned with any linear combination of $\Uhat$, $\mnvec_1$ or $\mnvec_2$. 

By solving the set of equations~\eqref{eq:nuAAp} one finds an expression for $\alpNP$ that is needed to yield a particular dataset 
$\left\{ \mnvec_{1}, \mnvec_{2}, \Uhat \right\}$,
\begin{align}
	\label{eq:alpNP}
	\alpNP 
	&=  \frac{   ( \mnvec_1 \times  \mnvec_2 ) \cdot \Uhat}
	{ ( \Uhat \times \vec{h}  ) \cdot  (  X_1\,\mnvec_2- X_2\,\mnvec_1 ) }\,,
\end{align}
assuming NP is the dominant contribution to nonlinearity. If linearity holds then $\alpNP\lesssim\sigma_{\alpNP} =  \sqrt{\Sigma_{k}(\partial \alpNP/\partial \CO_k)^2 \sigma_k^2}$. 
Hence, the sensitivity to probe $\alpNP$  is lost in the limit where the denominator in Eq.~\eqref{eq:alpNP} vanishes, because the NP contribution to nonlinearity is
\begin{align}
	\label{eq:NLNP}
	{\rm NL}_{\rm NP} = 
	\frac{\alpNP}{2}\,  ( \Uhat \times \vec{h} ) \cdot  (  X_1\,\mnvec_2- X_2\,\mnvec_1 ) \,.
\end{align}
It is straightforward to check that this happens under the conditions specified below Eq.~\eqref{KingNL}.

The presented method of limiting $\alpNP$, Eq.~\eqref{eq:alpNP}, contains theory input only in $X_i$ and $h_{AA'}$ which describe how NP affects the IS. The SM contribution in the factorized limit is fully parametrized by the observables $\vec \nu_i$ and $\vec\mu$. 
The form of $h_{AA'}$ depends on the assumed couplings of new physics to nuclei. For example, if the new interaction couples to quarks, then we expect that $h_{AA'}\propto AA'$~\cite{Haber:1978jt,Delaunay:2016brc}.  The atomic transition-dependent factors $X_{1,2}$ can be reasonably calculated by a many-body simulation (see the next section). 
This strategy is analogous to a search for NP, say, at the LHC, where all SM backgrounds are estimated using data driven methods and Monte Carlo simulation is used only in estimating the signal cross section. 

Thus far, most measurements of scalar-isotope shifts  
have been consistent with King linearity (see, however, the case of Samarium~\cite{SmExp:1981}). Nevertheless, some level of nonlinearity is expected to arise from SM higher-order contributions~\cite{0022-3700-15-7-009,PhysRev.188.1916,0022-3700-20-15-015,PhysRevA.31.2038}. 
These contributions, that are related to nuclear physics and electronic-structure dynamics linked together, are presently not understood in a quantitative manner for many-electron systems.  One possible source of nonlinearities is of the form of a field shift that depends on the isotope mass. Precision calculations recently showed that this effect is of $\CO(10^{-3}-10^{-4})$ in light atoms~\cite{Puchalski2010}. Likewise, such contributions in heavier elements with $Z=20-87$~\cite{0022-3700-20-15-015}, but only for $S\to P$ transitions, are estimated to be of a similar order. 
Hence, matching the precision of future measurements motivates the calculation of the remaining higher-order corrections.

If a deviation from King linearity is observed, it will be difficult to distinguish the NP and SM contributions to the nonlinearity. 
In this case there are two options in which further insight on NP can be obtained.
The first requires that the theory of King nonlinearity would advance and enable us to subtract the SM contributions, and in the process possibly gain new insight on the nature of nuclear effects in IS. To add to that, since nonlinearity in the case of NP is universal and in the case of SM specific to particular atomic configurations, a comparison between measurements in different systems will be beneficial. The second relies on the fact that NP forces are of longer range than nuclear effects which require overlap of the electronic wavefunction with the nucleus. Hence it might be possible to identify an observable that is less affected by the nucleus, but is still sensitive to the presence of long-range new physics interactions. In this regard, IS measurements involving Rydberg states might provide a smoking gun for the above types of NP.

For the proposed method to be effective, the element and the specific transitions should be chosen carefully. 
First, to make a significant progress as compared to current precision, we consider narrow optical clock transitions. The most accurate frequency measurements to date, with a relative error of $10^{-18}$ corresponding to sub-Hz accuracy, have been performed on narrow optical-clock transitions in laser-cooled atoms or ions~\cite{JunYe,NIST, NRC,Innsbruck,PTB,PhysRevLett.116.063001}.  Second, since the hyperfine interaction of electrons with the nucleus is a source for King nonlinearity~\cite{0022-3700-15-7-009}, we consider only even isotopes without nuclear spin. 

\section{Contribution of new bosons to Isotope Shifts} 
\label{sec:constNP}

In this section we discuss how theoretical IS predictions are modified in the presence of hypothetical new force carriers of spin $s=0,1$ or $2$ and mass $m_\phi$ which couple to electrons and neutrons with strength $y_e$ and $y_n$, respectively.
The effective spin-independent potential mediated by such bosons between the nucleus and its bound electrons is $V_\phi(r) =  -\alpNP (A-Z) e^{-m_\phi r}/r$, where $\alpNP=(-1)^sy_e y_n/4\pi$. Note that NP could also couple to protons, though without affecting the linearity of the King plot, hence we neglect such a coupling here.

To calculate the effect of this NP potential on atomic energies we use the ``finite field'' method where the potential is added directly to the Dirac equation in our many-body computations. The atomic structure calculations are variants of the combination of configuration interaction and many-body perturbation theory~(CI+MBPT)~\cite{dzuba96pra}. For the single-valence electron ions Ca$^+$ and Sr$^+$, we create an operator
$\hat\Sigma$ (see for example~\cite{Berengut:2003zg})
representing core-valence correlations to second order in the residual Coulomb interaction. This operator is added to the Dirac-Fock operator, along with the NP potential, to generate self-consistent solutions. In this approach, the sensitivity of a transition $i$ between electronic states $a$ and $b$ ($i=a\to b$) can be expressed
\be
	X_i = \frac{1}{A-Z} \left. \frac{d\epsilon_{ab} }{d\alpNP} \right|_{\alpNP =0} \, ,
\ee
%
where $\epsilon_{ab}$ is the 
difference of the energy levels of the states $a,~b$,
evaluated as a function of $\alpNP$ and the derivative is taken numerically at $\alpNP = 0$.

For neutral Sr, which has two valence electrons above closed shells, we use the CI+MBPT method as described in~\cite{Berengut:2005bz}. Briefly, we find the self-consistent solution of the Dirac-Fock equations, including the NP potential, for the closed-shell core (i.e. the $V^{N_e-2}$ potential where $N_e$ is the total number of electrons). 
In this potential we generate a set of B splines~\cite{PhysRevA.37.307,PhysRevLett.93.130405} which form a complete basis set. Valence-valence correlations are included to all orders using CI, while the core-valence correlations are included using second-order MBPT to modify the radial integrals.
The Yb$^+$ case is more complicated because of the hole transition, $4f^{14}\,6s \rightarrow 4f^{13}\,6s^2$. For this ion we use the particle-hole CI+MBPT method in the $V^{N_e-1}$ potential~\cite{berengut16pra} which has previously been used for Hg$^+$.

The many-body calculations can be cross-checked by perturbation theory, which yields
\begin{align}
	\label{eq:Xi}
	X_i = \int d^3r\frac{e^{-m_\phi r}}{r}\left[ |\Psi_b(r)|^2-|\Psi_a(r)|^2\right]\, , 
\end{align}
where $|\Psi(r)|^2$ is the electron-density evaluated in the absence of NP, and $h_{AA'}=AA'\,$amu for the  NP contribution in Eq.~\eqref{eq:nuAAp}.
As a cross-check of our many-body calculation, we use {\tt GRASP2K}~\cite{Jonsson20132197} to evaluate $|\Psi(r)|^2$  and compute $X_i$ using Eq.~\eqref{eq:Xi} for several Ca$^+$ transitions. We find good agreement between the two methods. 

We identify three regions of NP interaction range, separated by the electron wavefunction size, $a_0/(1+n_e)$, and the nuclear charge radius, $r_N\sim\,A^{1/3}\times(200\,$MeV$)^{-1}$. Here $a_0\approx(4\,$keV$)^{-1}$ is the Bohr radius and $n_e$ is the ionization number.
For $m_\phi \lesssim (1+n_e)/a_0$, the ``massless limit'', 
the interaction range is larger than the atomic size and $V_\phi\propto 1/r\,$ so that $X_i$ becomes independent of $m_\phi$. 
For intermediate masses, $(1+n_e)/a_0  \lesssim m_\phi \lesssim 1/r_N$,
the interaction range is within the size of the electron wavefunction, and the potential $V_\phi \propto e^{- m_\phi r}/r$ is mass-dependent. Hence, detailed knowledge of the electronic wavefunctions is necessary to evaluate the effect of NP.
In the heavy mass limit, $m_\phi \gtrsim 1/r_N$,
the interaction range is shorter than the nuclear radius and $V_\phi \propto \delta(r)/(m_\phi^2 r^2)$. In this limit, the NP and nuclear charge-radius effects are approximately aligned since $X_i \propto F_i \propto |\Psi_b(0)|^2-|\Psi_a(0)|^2$. This results in a suppressed sensitivity for new physics which scales  as $(X_{21} -F_{21}) \to 0$, see Eq.~\eqref{eq:modKing}, and~\cite{Delaunay:2016brc}.  

In the massless limit, $X_i$ can be estimated without a detailed computation of the atomic wavefunctions, as in this case the effective potential is Coulomb-like and thus its effects are approximately accounted for by a shift of $\alpha$, see Appendix~\ref{app:Ximassless}. 
We do not estimate the bounds on $\alpNP$ in the heavy mass limit as in this limit NP effects are indistinguishable from those of finite nuclear size. Bounds are therefore suppressed by a factor of $\CO(r_N/a_0 )$. 

\section{Current bounds and projections} 
\label{sec:Sensitivity}

Here we derive the constraints on the product of electron and neutron coupling, $y_e y_n$, from existing IS data of Ca$^+$  and project the bounds for different transitions alkali-like systems in the 10\,eV-50\,MeV mass range, assuming that better IS data will be available in the future. Our results are summarized in Fig.~\ref{fig:bounds} and Tab.~\ref{tab:sensitivity}.

\subsection{Constraints from King linearity}
\label{sec:current}

We apply our method to available IS data of Ca$^+$ (solid line of Fig.~\ref{fig:bounds}). 
In the massless-boson limit, $m_\phi\lesssim10\,$keV, the bound is essentially independent of $m_\phi$. 
 At the high mass limit, we expect that $F_{21} = X_{21}$. Since the theoretical control of $F_{21}$ is worse than the experimental error, one can get an incorrect $m_\phi$ dependence of the $y_n y_e$ bound at that limit. However, the ratio $F_{21}^{\rm th} /X_{21}$ ($F^{\rm th}_{21}$ is the theoretical value calculated in the absence of NP) has much smaller error. Thus, in order to account for the reduction in sensitivity as $m_\phi$ increases, we rescale the $y_e y_n$ bound by  $(1- F^{\rm exp}_{21}/ X_{21} )/(1 - F_{21}^{\rm th} /X_{21})$, where $F_{21}^{\rm exp}$ is the measured value of $F_{21}\,$.
We verified with {\tt GRASP2K} that this factor does not change by more than a few percent if the charge radius is changed by order one (that is known to a few percent accuracy, hence this is a rather conservative approach).
Indeed we see that for $ m_{\phi}>  Z \alpha m_e$ the limits get weak, and the sensitivity decreases approximately as $ m^{-3}_{\phi}$ for large masses.
In Appendix~\ref{app:high-mass} we give two heuristic arguments that obtain this asymptotic scaling of our loss of sensitivity\footnote{We thank Richard Hill and Clara Peset for an enlightening conversation}: the first is based on approximating Eq.~\eqref{eq:Xi}; and the second is based on a non-relativistic QED (NRQED) effective theory approach.

For current bounds, we consider Ca$^+$ ($Z=20$).
There are five zero-nuclear-spin, stable or long-lived isotopes with $A=40, 42, 44, 46, 48$.
Refs.~\cite{PhysRevLett.115.053003,Shi2016} reported IS measurements for three isotope pairs ($A=42, 44, 48$ relative to 40) 
in three dipole-allowed transitions in Ca$^+$ at wavelengths of 397.0\,nm ($S\to P$), 866.5\,nm ($D\to P$) and 854\,nm ($D\to P$, not used here)  with an uncertainty of $\CO(100)\khz$. 
 
\begin{table}[t]
\begin{center}
\begin{tabular}{|c|c|c|c|c|c|}
\hline\hline 
& transition 1 & transition 2 &  accuracy  & $y_e y_n$ bound \\
& [nm] & [nm] & $\sigma_i$ & ($m_\phi=0$) \\
\hline\hline 
Ca$^+$ & 397.0& 866.5& $0.1\,{\rm MHz}$ & $2\cdot 10^{-9}$ \\
\hline\hline
Ca$^+$  & 729.3 & 732.6 & $1\,{\rm Hz}$ & $2\cdot10^{-14}$ \\
\hline
Sr$^+$  & 674.0 & 687.0 & $1\,{\rm Hz}$ &  $2\cdot 10^{-13}$ \\
Sr/Sr$^+$  & 698.4 & 674.0 & $1\,{\rm Hz}$ & $3\cdot 10^{-15}$ \\
\hline
Yb$^+$ & 435.5 & 466.9& $1\,{\rm Hz}$ & $2\cdot 10^{-15}$ \\
\hline\hline
\end{tabular}
\caption{
The 95\,\%\,CL bounds on $y_e y_n$ for a massless mediator $\phi$ from Ca$^+$  
data~\cite{PhysRevLett.115.053003}
and 95\,\%\,CL projections for Ca$^+$, Sr$^+$, Sr/Sr$^+$ and Yb$^+$ assuming on error of $\sigma_i=1\,$Hz. 
}
\label{tab:sensitivity}
\end{center}
\end{table}%

Among the non-IS experiments that probe $y_e y_n-m_\phi$ parameter space,
we consider here only the ones most sensitive to new light bosons coupled to electrons and neutrons. The shaded regions in Fig.~\ref{fig:bounds} summarize the current reach of these experiments.
We stress, however, that some of them are derived involving of further theory assumptions, in contrast to our method which relies on few theory inputs. 
For new bosons lighter than few$\times100\,$eV, fifth force experiments~\cite{Bordag:2001qi,bordag2009advances} are potentially sensitive. 
Since the interaction range covered by these experiments is much larger than the atomic size, only forces with non-zero atomic coupling can be probed. For illustration we show in Fig.~\ref{fig:bounds} the fifth force bound applicable to U(1)$_{\rm B-L}$ gauge bosons~\cite{Harnik:2012ni}.

Furthermore, separately $y_n$ is constrained by various neutron scattering experiments~\cite{Leeb:1992qf,Nesvizhevsky:2007by,Barbieri:1975xy}  and 
 $y_e$ by the anomalous magnetic moment of the electron  $(g-2)_e$~\cite{Olive:2016xmw,Hanneke:2010au}  and by electron beam-dump experiments for $m_{\phi} > 1$\,MeV.

Both $y_e$ and $y_n$ are also severely constrained by globular cluster energy loss for masses $ m_{\phi} \lesssim 10$ keV \cite{Raffelt:2012sp,Yao:2006px,Grifols:1986fc,Grifols:1988fv} down to $y_e y_n<10^{-25}$ and $y_e$ by sun cooling~\cite{Redondo:2008aa,Jaeckel:2010ni}. Couplings to nucleons in the $ 10^{-10}-10^{-7}$   range for $m_{\phi} \lesssim 100$ MeV may be also excluded by  energy loss in the core of SN1987A \cite{Raffelt:2012sp,Blum:2016afe}.
However, such  astrophysical bounds might be avoided in certain models such as chameleon, see \cite{Burrage:2016bwy} and references therein. 
In order to derive an upper bound on $y_e y_n$, we combine for each mass the best constraint on $y_n$ from neutron experiments with $y_e$ either from $(g-2)_e$ or from astrophysics.

\subsection{Prospect for future measurements}
\label{sec:future}

As the precision of optical spectroscopy continues to improve, higher accuracy IS measurements in different systems can be achieved in the near future. Accordingly, we estimate the sensitivity that would be achieved for several transitions in alkali or alkali-earth ions or atoms, given the improved accuracy.  

Here we consider a comparison between the two fine-structure split electric quadrupole transitions in Ca$^+$ and Sr$^+$. 
A comparison between the optical clock transitions in Sr$^+$ and Sr, and the quadrupole and octupole transitions in Yb$^+$ are also presented. In principle, to enhance the sensitivity of our method, it is desired to compare transitions that involve levels that are as different as possible. For this reason comparing the two fine-structure split electric quadrupole transitions in Ca$^+$ or Sr$^+$ is not ideal, especially when compared to the sensitivity of the E2 and E3 lines in Yb$^+$ or comparing the E2 line in Sr$^+$ with the intercombination line in Sr. We include these transitions in our projections since their high-resolution IS measurement is experimentally simpler.  

All the transitions above are expected to be measured with $1$\,Hz accuracy. 
Under the assumption that King linearity will hold in those future measurements and following Appendix~\ref{app:alpNPproj}, the projected bounds are plotted in Fig.~\ref{fig:bounds}~(dashed lines) as a function of $m_\phi$ and summarized in the lower part of Tab.~\ref{tab:sensitivity}. 
The resonance structures, around the 10~keV scale, arise from cancellations in the denominator of Eq.~(\ref{eq:alpNP}). These local losses of sensitivity at different masses per atomic system provide another motivation for IS measurements in complementary systems for a good coverage of the parameter space.

The various projections with 1\,Hz accuracy significantly improve the bounds in the $m_\phi\geq 10\kev$ region in parameter space. For lower $m_\phi$ they are weaker than astrophysical bounds. However, astrophysical bounds are subject to large uncertainties and can be broken by models such as the chameleon effect~\cite{Burrage:2016bwy}. Thus, an independent laboratory bound in this low-mass region is nevertheless worthwhile. For $m_\phi\sim$\,a\,few\,MeV the projections of Ca$^+$ ($S\to D$ transitions) and Sr$^+$ are comparable to $y_ey_n$ from neutron scattering~\cite{Leeb:1992qf} and $(g-2)_e$. Since neutron experiments are affected by uncertainties~\cite{Wissmann:1998ta,Nesvizhevsky:2007by,Antoniadis:2011zza,TuckerSmith:2010ra} such as those in electron-neutron scattering length, nuclear input values and missing higher-order terms in the neutron-scattering cross section, the bounds in the high-mass range well above the neutron energies of $E_n<10\kev$~\cite{Leeb:1992qf} should be understood as an indication of the order of magnitude. Consequently, theoretically cleaner IS probes at the same order will already improve the bound robustness. 
Note that a Sr/Sr$^+$ (Yb$^+$) IS comparison would become more effective than other existing methods in probing new bosons above $\sim10\kev$ already with $100\,$Hz ($1\,$kHz) accuracy (the bound related to Sr/Sr$^+$ constructed from comparison of transition involving neutral and ion systems suffers from some numerical instabilities for masses above $20\,$MeV and thus is not shown).
Finally, let us note the range of $y_ey_n$ needed to explain the Be anomaly~\cite{Feng:2016jff,Feng:2016ysn} can be probed by future IS measurements of Yb$^+$  at the 1\,Hz level. 


\section*{Acknowledgment} 
\label{sec:ack}

We thank J.~Feng, O.~Firstenberg, I.~Galon, R.~Hill, S.~G.~Karshenboim, K.~Pachucki, C.~Peset, M.~Pospelov, P.~O.~Schmidt and T.~Tait for fruitful discussions. CD and YS are particularly thankful to J.~Ekman and P.~J$\ddot{{\rm o}}$nsson for their help regarding the {\tt GRASP2K} program.
The work of DB is supported by the DFG Reinhart Koselleck project, and the ERC (Dark-OST Advanced Project).
The work of CD is suported by the ``Investissements d'avenir, Labex ENIGMASS''. 
The work of CG is supported by the European Commission through the Marie Curie Career Integration Grant 631962 and by the Helmholtz Association through the recruitment initiative programme.
The work of RO is supported by grants from the ISF, I-Core, ERC, Minerva and the Crown photonics center.
The work of GP is supported by grants from the BSF, ERC, ISF, Minerva, and the Weizmann-UK Making Connections Programme.
The work of YS is supported by the U.S. Department of Energy~(DOE) under grant contract numbers DE-SC-00012567 and DE-SC-00015476.

\bibliographystyle{utphys}
\bibliography{IS_light_bib}


\clearpage
\newpage
\maketitle
\onecolumngrid
\begin{center}
\textbf{\large Constraining new light force-mediators by isotope shift spectroscopy } \\ 
\vspace{0.05in}
{ \it \large Supplementary Material}\\ 
\vspace{0.05in}
{Julian~C.~Berengut, Dimtry~Budker, C\'edric~Delaunay, Victor~V.~Flambaum, Claudia~Frugiuele, 
Elina~Fuchs, Christophe~Grojean, Roni~Harnik, Roee~Ozeri, Gilad~Perez, and Yotam~Soreq}
\end{center}
\onecolumngrid
\setcounter{equation}{0}
\setcounter{figure}{0}
\setcounter{table}{0}
\setcounter{section}{0}
\setcounter{page}{1}
\makeatletter
\renewcommand{\theequation}{S\arabic{equation}}
\renewcommand{\thefigure}{S\arabic{figure}}
\renewcommand{\thetable}{S\arabic{table}}
\newcommand\ptwiddle[1]{\mathord{\mathop{#1}\limits^{\scriptscriptstyle(\sim)}}}

\section{Visualizing the vector space}
\label{app:geometry}

\begin{figure*}[b]
\begin{center}
\includegraphics[width=8.5cm]{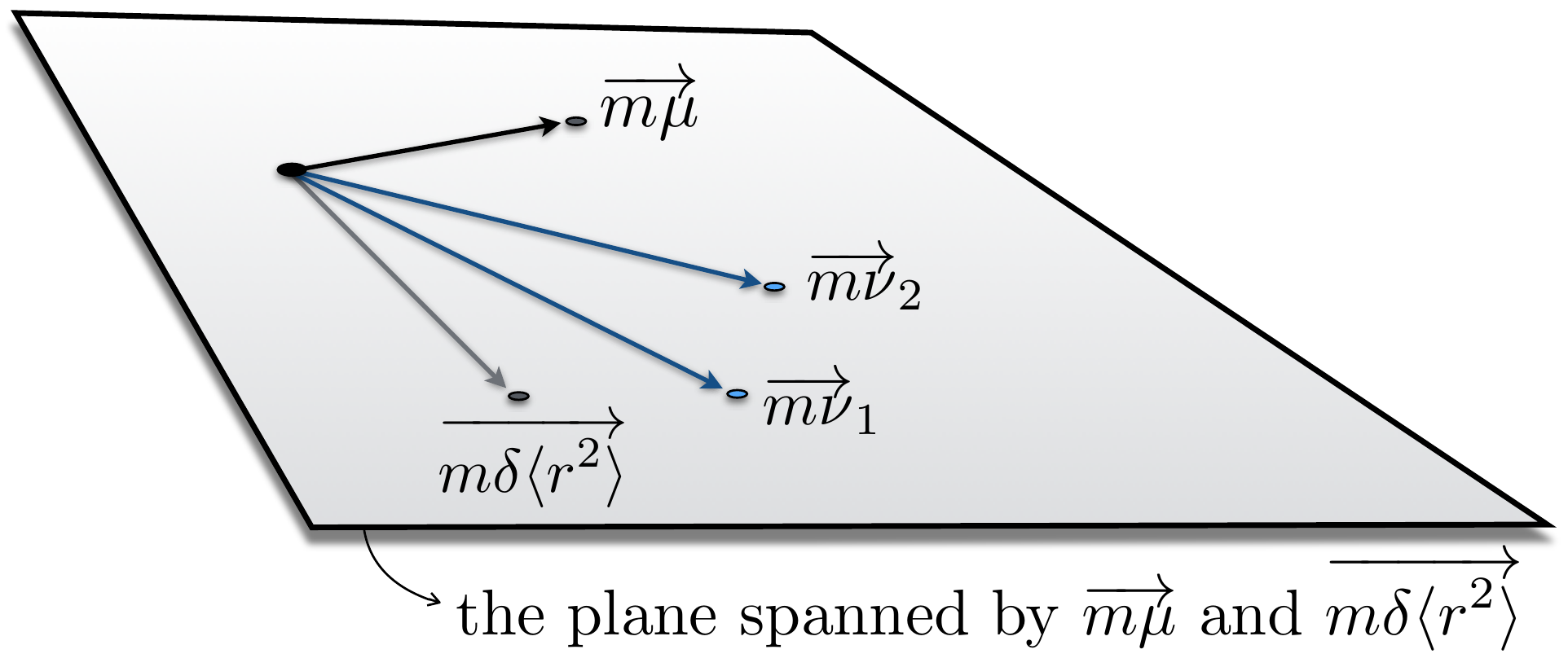}
\includegraphics[width=8.5cm]{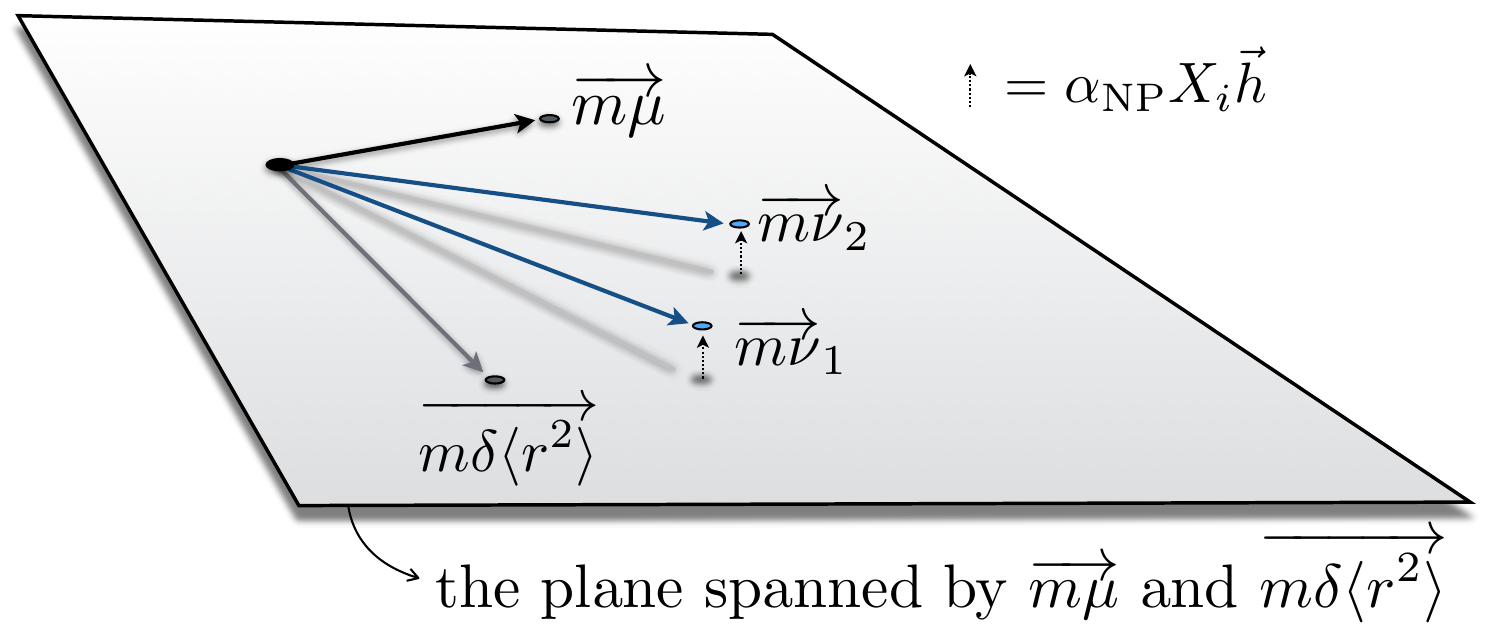}
\caption{Left: A cartoon of the prediction of factorization, Eq.~\eqref{eq:nuAApVec} in vector language. All of the isotope shift measurements (which are here three dimensional vectors $\protect\mnvec_{1,2}$) lie in the plane that is spanned by $\protect\Uhat$ and $\protect\mrvec$. This coplanarity can be tested by measuring whether $\protect\mnvec_1$, $\protect\mnvec_2$ and $\protect\Uhat$ are coplanar. Right: In the presence of new physics the isotope shift get a contribution which can point out of the plane. A new long range force can spoil the coplanarity of $\protect\mnvec_1$, $\protect\mnvec_2$ and $\protect\Uhat$. 
}\label{fig:plane}
\end{center}
\end{figure*}

\begin{figure}[b]
\begin{center}
\includegraphics[width=7cm]{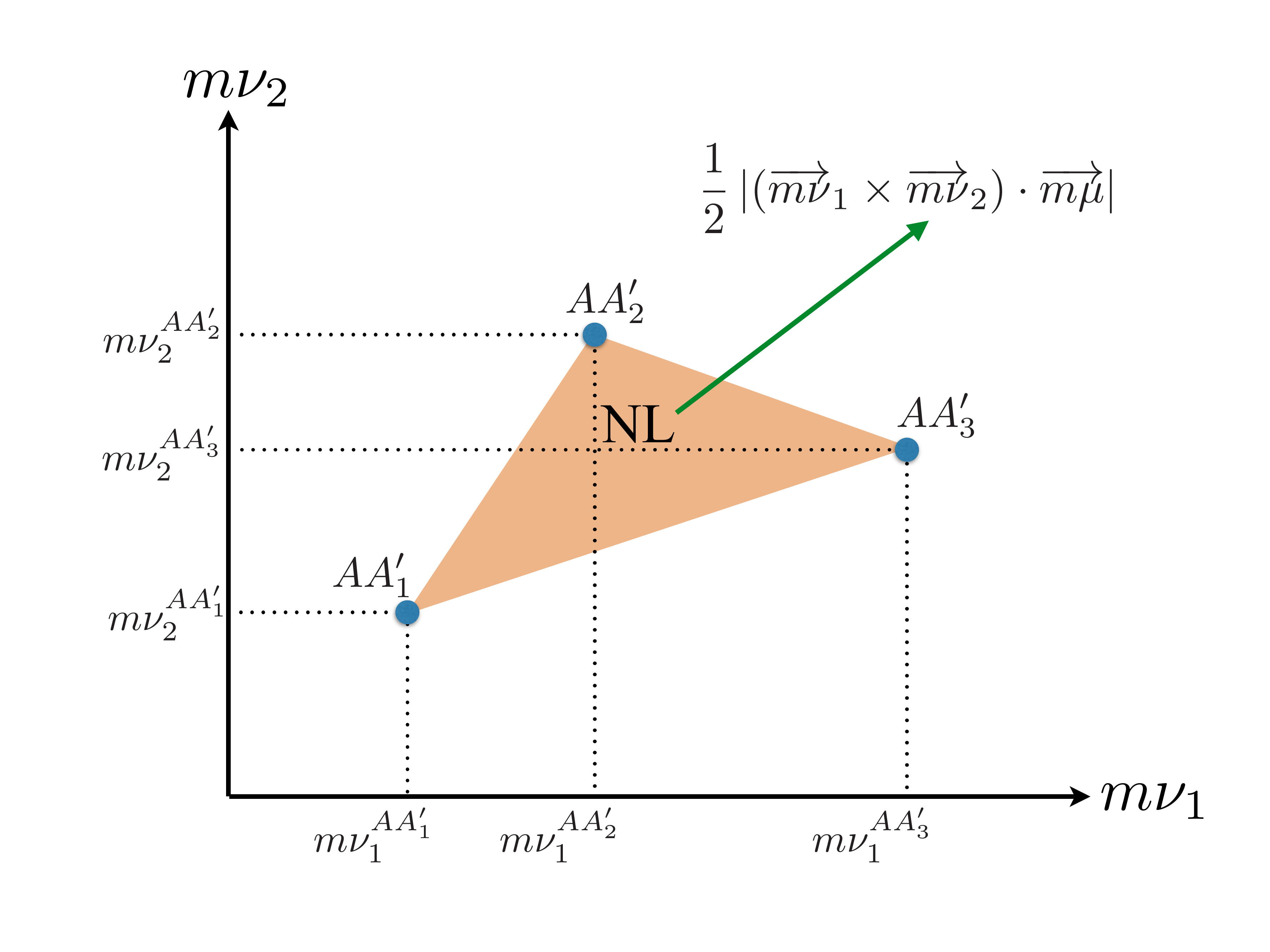}
\caption{
Illustration of nonlinearity in the King plot of the isotope shifts $\protect\mnvec_{1,2}$, as defined in Eq.~\eqref{eq:mnuvec}, in isotope pairs ${A A'_j}, j=1,2,3$. 
The area of the triangle corresponds to the NL of Eq.~\eqref{eq:NL}. 
}
\label{fig:NL}
\end{center}
\end{figure}

In the main text we define the following vectors in the $A'$ vector space 
\begin{align}
	\mnvec_i &\equiv \left( m \nu_i^{AA'_1},m \nu_i^{AA'_2}, m\nu_i^{AA'_3} \right)\,, \\
	\mrvec &\equiv \left( \langle r^2 \rangle_{AA'_1}/\mu_{AA'_1}, \langle r^2 \rangle_{AA'_2}/\mu_{AA'_2}, \langle r^2 \rangle_{AA'_3}/\mu_{AA'_3} \right) \, , \\
	\Uhat &\equiv (1,1,1) \, .
\end{align}
As long as $\mnvec_{1,2}$ are spanned by  $\Uhat$ and $\mrvec$, the resulting King plot will be linear. 
In Fig.~\ref{fig:plane}, we illustrate the vector space of the various components related to isotope shifts that leads to the nonlinearites. The NP contribution to IS, $\alpNP X_i \vec{h}$, may lift the IS vectors from the $(\Uhat,\mrvec)$ plane, resulting in a nonlinear King plot.    
Fig.~\ref{fig:NL} illustrates a nonlinear King plot, where the area of the triangle corresponds to the NL of Eq.~\eqref{eq:NL}. 

\section{Derivation of $X_i$ in the $m_\phi\to0$ limit}
\label{app:Ximassless}

Here we estimate the NP contribution $X_i$ to IS in the special case where the force-carrier is much lighter than the inverse atomic size, $m_\phi \ll (1+n_e)/a_0\sim\mathcal{O}($few$\,$keV). Since in this limit the effective potential is Coulomb-like, $V_\phi(r) \simeq (A-Z) \alpNP/r$, $X_i$ can be simply estimated through a shift of the fine-structure constant $\alpha$, without a detailed calculation of the electronic wavefunctions. At fundamental level, the Coulomb potential is modified by the shift
\beq\label{eq:Fundshift}
\alpha Z \to \alpha Z + \alpNP (A-Z)\,.
\eeq
In the absence of NP, the binding energy of the atomic level $a$ for isotope $A$ scales as 
\begin{align}
	\label{eq:EAa}
	E^A_{a} = (\alpha Z^{a}_{\rm eff})^2 I^A_{a}\,,
\end{align}
where $Z^{a}_{\rm eff}\equiv Z -\sigma_a$ is the effective nuclear charge seen by the valence electron in the state $a$, and $I^A_{a}$ is a constant independent of the charge (modulo $\CO(\alpha Z_{\rm eff}^a)^4$ corrections from the fine-structure). The constant $\sigma_a>0$ accounts for the screening due to inner electrons. A similar screening effect may occur for the new physics force such that Eq.~\ref{eq:Fundshift} implies to shift the physical observables as 
\begin{align}
	\alpha Z_{\rm eff}^a \to\alpha Z_{\rm eff}^a+ \alpNP (A-Z-\sigma'_a)\,,
\end{align}
where the constant $\sigma'_a$ accounts for the screening of the nuclear NP charge by inner electrons.  Note that precise knowledge of this constant is not crucial since it is universal (as a first approximation) for all isotopes and will therefore cancel in $X_i$. 

Hence the prediction for the IS of the transition $i=a\to b$ is 
(in natural units, i.e.~the reduced Planck constant is set to $\hbar=1$)
\begin{align}
	\nu^{AA'}_i 
=& 	( E^A_{b} - E^A_{a}) - ( E^{A'}_{b} - E^{A'}_{a})  \nonumber\\
=&	\left[ \alpha Z^b_{\rm eff}+ \alpNP(A-Z-\sigma'_b) ]^2 I^A_{b} - [ \alpha Z^a_{\rm eff}+  \alpNP(A-Z-\sigma'_a)^2]  I^A_{a}   \right] 
	- (A\to A') \,,
\end{align}
and expanding to leading order in $\alpNP \ll \alpha$ yields
\begin{align}
	\nu^{AA'}_i 
\approx 	\nu^{AA'}_i\big|_{\alpNP=0} 
	+\frac{2\alpNP}{\alpha} \left[  A  \left(\frac{E^A_{b}}{Z_{\rm eff}^b}  -  \frac{E^A_{a}}{Z^a_{\rm eff}}\right) - (A\to A')\right]  \, .
\end{align}
Since IS are typically orders of magnitude smaller than the transition frequencies (with the exceptions of very degenerate states such as in dysprosium~\cite{PhysRevA.50.132,Leefer:2014tga}), we can take $E^{A'} \approx E^A$  in the NP contribution above and matching to Eq.~\eqref{eq:nuAAp} we find $\gamma_{AA'}=A-A'$ and
\begin{align}\label{eq:Ximassless}
	\left. X_i \right|_{m_\phi=0} 
	\approx 
	2\alpha^{-1}\left( \frac{E_{b} }{ Z_{\rm eff}^b} - \frac{E_a }{ Z_{\rm eff}^a} \right)  \, .
\end{align}
By a combination with Eq.~\eqref{eq:alpNP}, the above expression provides a reasonable estimate of the constraint on $\alpNP$ without the need for an accurate knowledge of electronic densities. For example, we found that the upper bounds on $y_ey_n$ obtained by evaluating $X_i$ with Eq.~\eqref{eq:Ximassless} and with the CI+MBPT method are comparable and only differ by $\CO(1)$ factors.

\section{Projecting future bounds}
\label{app:alpNPproj}

Our procedure above applies to cases with enough experimental data. For systems lacking  (sufficiently precise) measurements, we can still derive projections provided that an acceptable estimation of the $F_{21}$ constant is available from either theory calculation or hyperfine splitting data (whenever available).
Assuming the observation of linearity and global experimental uncertainties $\sigma_{i}$ of $\nu_i^{AA'}$, the only missing information is how much the new physics vector, $\vec{h}$, points towards the linearity plane. 
While a precise determination of the vector component requires data, the projection of the NP vector  $h_{AA'}=AA'\,$amu along the mass shift direction is easily obtained without the need of experimental input.
Therefore, a best-case projection\footnote{The actual bound obtained by data will be always weaker, $ [\sigma\alpha_{\rm NP}]_{\text{proj}} \leq [\sigma\alpha_{\rm NP}]_{\text{data}} $ since the projection neglects the alignment with the FS that will weaken the bound.}
$[\sigma_{\alpha_{\rm NP}}]_{\text{proj}} $ can be obtained by neglecting the possible additional alignment of NP with nuclear effects,
\begin{align}
	\label{eq:delalpNP}
	[\sigma_{\alpha_{\rm NP}}]_{\text{proj}}	\!\ \sim \! 
	  \frac{ \sqrt{\sigma^2_2 + \sigma^2_1F_{21}^2}  }{ \left(  X_2  - X_1 F_{21} \right)}
	\frac{ A }{ \Delta A_j^{\rm min} \Delta A_j^{\rm max}   } \, ,	
\end{align}
where $\Delta A_j^{\rm min(max)}\equiv{\rm min\,(max)}[A-A_j] $ and $\sigma_i$ is the assumed standard deviation of IS measurements in transition $i$. Note that the sensitivity to $\alpNP$ is weaker by a factor of $A/\Delta A_j^{\rm min}$ than the naive expectation of Refs.~\cite{Delaunay:2016brc,Frugiuele:2016rii,Delaunay:2016zmu}. The reason is that the NP physics vector lies mostly in the linearity plane, in particular it has a large projection along the mass shift direction.
Eq.~\eqref{eq:delalpNP} implies that elements with small $A/(\Delta A_j^{\rm min} \Delta A_j^{\rm max})$ are preferred. However, if the mass shift dominates over the field shift, which is the case for light elements, the sensitivity is reduced as well. 

\section{Scaling of High Mass Limits}
\label{app:high-mass}

As we discussed in Section~\ref{sec:NP}, King linearity (or coplanarity) is only sensitive as a probe of long-range forces, which extend beyond the nuclear size. This is because the contribution of a short-range interaction to the isotope shift can be absorbed into the contact interactions such as the nuclear charge radius. From the perspective of an atom, a new interaction begins to approach a contact interaction when its range is shorter than the effective radius of the inner most K-shell electron, $(Z\alpha m_e)^{-1}=a_0/Z$. We thus expect our method to start losing power for mediators heavier than $Z\alpha m_e$ as can be seen in Fig.~\ref{fig:bounds}. 

It is instructive to investigate like what power of $m_\phi/(Z\alpha m_e)$ one would expect the limits to decrease.  As shown in Eq.~\eqref{KingNL} and the surrounding discussion, the contribution of NP to nonlinearity is proportional to $(X_2 - X_1 F_{21} )$. When the NP contribution is aligned with the FS, $X_i\propto F_i$, this factor vanishes. We must thus investigate how $X_i$ and $F_i$ differ for the various transitions in the limit of $m_\phi\rightarrow \infty$. In our proposed procedure, $F_i$ is measured from data and $X_i$ is estimated from theory, as described in Section~\ref{sec:constNP}, leading to the limits shown in Fig.~\ref{fig:bounds}. Since these two quantities are extracted using different methods we would like to ensure that the alignment of $X_i$ and $F_i$ is captured correctly and that the weakening of the bounds for $m_\phi > Z\alpha m_e$ agrees with our theoretical expectation.

The asymptotic behavior of the limits in Fig.~\ref{fig:bounds} exhibits an approximate $m_\phi^3$ behavior. We will now show that this may be expected on theoretical grounds, first in position space, using a hydrogen-like approximation and then in momentum space, using an effective field theory~(EFT). 

To leading order in the small new physics coupling, the contribution of a new Yukawa potential to the energy of the atomic level $a$ is
\begin{equation}
	\Delta E_a = \alpNP \int d^3 r  \frac{e^{-m_\phi r}}{r} \left|\Psi_a (r) \right|^2 \, .
\end{equation}
To study the scaling behavior in a simple case, we approximate the multi-electron crudely as a single-electron atom with an effective $Z^{(a)}_\mathrm{eff}$ which will account for the screening of the nucleus by the inner shells. Of course, the full calculation of Section~\ref{sec:constNP} accounts for the multi-electron effects fully and the approximation we use here should be taken as a toy model to study the scaling of our bounds.  We consider atoms in the $n$S state, with ($n=1,2,\ldots$), which is relevant for us since many of our isotope shifts involve an S state. Using the explicit form of non-relativistic hydrogen-like wavefunctions, it is simple to show that in this case
\begin{equation}
	\label{eq:deltaEa}
	\Delta E_{n\mathrm{S}} 
	= \alpNP \frac{\left|\psi_{n\mathrm{S}} (0) \right|^2}{m_\phi^2} \left( 1- \frac{4Z^{(n\mathrm{S})}_\mathrm{eff}\alpha\, m_e}{m_\phi} + \ldots \right) \, ,
\end{equation}
where we assumed that $m_\phi \gg Z \alpha m_e$ and expanded in powers of $1/m_\phi$, keeping the first two terms. The leading term is to be expected, and is identical to the shift in energy from a contact interaction potential $\delta^3(r)/m_{\phi}^2$. This term will thus lead to isotope shifts in the $a\to b$ transition that are proportional to $|\Psi_a(0)|^2-|\Psi_b(0)|^2$ and thus proportional to the FS. This term therefore cancels in the combination $(X_2 - X_1 F_{21} )$. 
The second term in Eq.~(\ref{eq:deltaEa}) has an additional factor of $Z_\mathrm{eff}$ leading to an IS proportional to $Z_\mathrm{eff}^{(a)} |\Psi_a(0)|^2 - Z_\mathrm{eff}^{(b)}|\Psi_b(0)|^2$. 
This term will not cancel in $(X_2 - X_1 F_{21})$, and since it scales as $m_\phi^{-3}$ we would expect the bounds to lose power as $m_\phi^{3}$ at high mass\footnote{It is interesting to note, that for a single electron atom, where $Z_\mathrm{eff}=Z$ for all states, the isotope shift again scales as $|\Psi_a(0)|^2-|\Psi_b(0)|^2$. In this case the bound will scale as $m_\phi^{4}$ from the higher order terms that were dropped in Eq.~(\ref{eq:deltaEa}).}. 

We can also understand the scaling of our bounds at high mediator mass in an EFT, where the heavy mediator was integrated out. The appropriate EFT for this purpose is known as non-relativistic QED~(NRQED)~\cite{Caswell:1985ui}. This EFT is often used to study nuclear effects in atoms, but here we will use it to describe our heavy mediator. In this theory the effects of the heavy mediator are captured by four-fermion interactions suppressed by powers of $m_\phi$. 
The reader who is familiar with NRQED may be surprised by the $m_\phi^{-3}$ because the theory contains operators that are suppressed  by $m_\phi^{-2}$ and by $m_\phi^{-4}$ but none that are suppressed by a cubic power~\cite{Hill:2012rh}. This however can be resolved within NRQED as we now show, since the $m_\phi^{-3}$ is simply a 1-loop correction to the $m_\phi^{-2}$ Wilson coefficient\footnote{We thank Richard Hill for an enlightening conversation.}.

To demonstrate the $m_\phi^{-3}$ scaling one needs to do a 1-loop matching between the full Yukawa theory and the EFT. This calculation amounts to calculating the LO QED correction to the tree level calculation in the Yukawa theory. 
To perform the matching we calculate a physical quantity twice, once in the full theory and once in the EFT, and then set the Wilson coefficient in the EFT to get a similar result order by order in perturbation theory. 
For the full theory we take non-relativistic QED with an additional Yukawa interaction. For simplicity we take the nucleus to be infinitely heavy and of zero size. The nuclear charge will again be set to $Z_\mathrm{eff}$, in accordance with the approximation taken above for electron screening effects.
The Yukawa propagator in momentum space is $(q^2 + m_\phi^2)^{-1}$ and the rest of the Feynman rules are shown in~\cite{Kinoshita:1995mt}.

The EFT we will match onto is NRQED, again with an infinitely heavy nucleus, but with the inclusion of one contact interaction between the nucleus, $N$, and the electron $C (\psi_e^*\psi_e)(\psi_N^*\psi_N)/m_\phi^2$, where $C$ is a Wilson coefficient. 
Since we are only interested in the scaling behavior we will not dwell on precise numerical coefficients, but will focus on the parametric scaling.
A quantity that is simple to compute for this matching calculation is the two-electron correlation function setting external momenta to zero, thus avoiding complications of bound states.  

\begin{figure}
\begin{center}
\includegraphics[width=7.5cm]{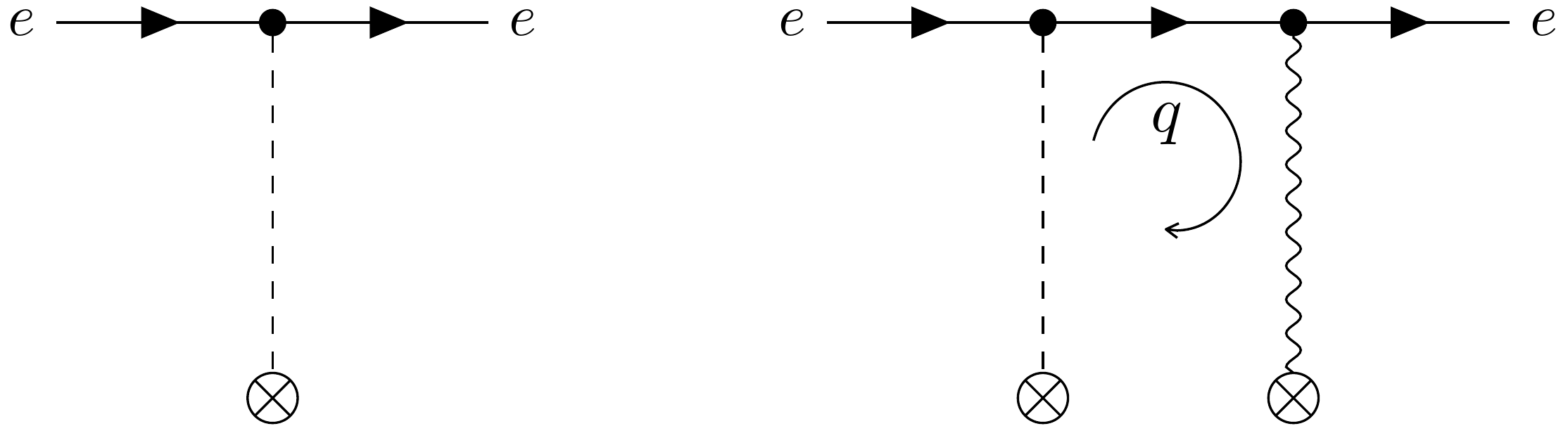} 
\caption{Feynman diagrams for the two point function of the electron in the presence of a nucleus. We take external momenta to zero. The tree level diagram is  the first correction of the Yukawa potential and the loop calculation is the first QED correction to it. The position-space calculation uses the full hydrogen wavefunction and thus effectively re-sums all such corrections with an arbitrary number of photon exchanges.   }\label{fig:matching}
\end{center}
\end{figure}

We begin by matching at tree level. In the full theory, to LO in the NP interaction, the tree-level contribution to the 2-point function is shown in Fig.~\ref{fig:matching} and is trivially $ \alpNP/m_\phi^2$.
In the EFT, a similarly simple calculation produces $C/m_\phi^2$ for the two-point function. As expected, at tree level the matching gives
\begin{equation}
\label{eq:treeMatching}
C_\mathrm{tree} = \alpNP
\end{equation}

Still at LO in NP, we can calculate to an additional order in the QED coupling  $\alpha$ which arrises at one loop and is also shown in Fig.~\ref{fig:matching}. 
Since Coulomb scattering diverges as the incoming velocity goes to zero, we will not be surprised to encounter an IR divergence in this calculation. However, as expected, the IR divergence has an identical structure in the full and effective theories and will thus not affect the value of the Wilson coefficient at order $\alpha$.  
In the full theory we find the two point function
\begin{equation} 
	\label{eq:1loopfull}
	\mbox{Full theory:}\qquad
	D_{ee}^{(1)}=
	\int \frac{d^3 q}{(2\pi)^3}\, \frac{\alpNP}{q^2+m_\phi^2}\, \frac{1}{E-\frac{q^2}{2 m_e}}\,
	\frac{Z_\mathrm{eff} e^2}{q^2+\lambda^2} 
	\sim -\frac{2 \alpNP Z_\mathrm{eff}  \alpha m_e}{\lambda m_\phi^2} + \frac{2 \alpNP\, Z_\mathrm{eff}\alpha m_e}{m_\phi^3} +\ldots
\end{equation}
where the three terms in the integrand are the Yukawa, electron and Coulomb propagator,  respectively, and we used $\alpha=e^2/4\pi$ and E=0. The coulomb potential is IR-regulated by $\lambda$.
The result in Eq.~\eqref{eq:1loopfull} has been expanded at large $m_\phi$. The second term in Eq.~\eqref{eq:1loopfull} is already reminiscent of the $m_\phi^{-3}$ term in Eq.~(\ref{eq:deltaEa}). Repeating the $O(\alpha)$ computation in the EFT is straightforward, 
\begin{equation} 
	\label{eq:1loopEFT}
	\mbox{EFT:}\qquad
	D_{ee}^{(1)}=\qquad\int \frac{d^3 q}{(2\pi)^3}\, \frac{C}{m^2_\phi}\, \frac{1}{E-\frac{q^2}{2 m_e}}\,\frac{Z_\mathrm{eff} e^2}{q^2+\lambda^2} \sim -\frac{2 C Z_\mathrm{eff}  \alpha m_e}{\lambda m_\phi^2} \,.
\end{equation}
As expected, using the tree-level value for $C$, Eq.~(\ref{eq:treeMatching}), the IR divergences in the full and effective theories match at order $\alpha$. However,  the second term in Equation~(\ref{eq:1loopfull}), leads to a finite correction to the Wilson coefficient at $O(\alpha)$, 
\begin{equation}
\label{eq:wilson}
\Delta C = \frac{2\alpNP Z_\mathrm{eff}\alpha m_e}{m_\phi}=\frac{2\alpNP Z_\mathrm{eff}}{a_0 m_\phi}\,.   
\end{equation}

We thus find that the 1-loop correction to the $m_\phi^{-2}$ Wilson coefficient shown in Fig.~\ref{fig:matching} leads to an effective operator which scales as $Z_\mathrm{eff}/(a_0 {m_\phi})^3$, in qualitative agreement with our position-space calculation. The two calculations are related. The position space calculation makes use of the hydrogen wavefunction and thus in some sense re-sums diagrams with an arbitrary number of photon exchanges between the nucleus and the electron before and after the scalar line. The EFT calculation isolates the first of these corrections. In both calculations we have accounted for multi-electron effects by taking the nuclear charge to be an effective one, with the potential being Coulombic otherwise. Again, we find that the mismatch between the $Z_\mathrm{eff}$'s does not allow this term to be absorbed in the FS.
We are, however, not surprised that the numerical coefficients of the $m_\phi^{-3}$ terms in Eqs.~\eqref{eq:deltaEa} and~\eqref{eq:wilson} do not agree since the former captures bound state dynamics, properly resumming IR effects that are dominant at the Bohr radius. 
 
\end{document}